\documentclass[12pt]{article}
\usepackage{amsmath}
\usepackage{amssymb}
\usepackage{graphics}
\usepackage{epsf,graphicx,rotating,color}
\usepackage{bm,bbm}
\usepackage{amsfonts}

\begin{document}
\title{{\bf  Diamond as a solid state quantum computer with a linear chain of nuclear spins system}}
\author{ G.V.  L\'opez\footnote{gulopez@udgserv.cencar.udg.mx}\\\\Departamento de F\'{\i}sica de la Universidad de Guadalajara,\\
Blvd. Marcelino Garc\'{\i}a Barrag\'an 1421, esq. Calzada Ol\'{\i}mpica,\\
44430 Guadalajara, Jalisco, M\'exico}
\date{September, 2013}
\maketitle

\begin{abstract}
By removing a $^{12}C$ atom from the tetrahedral configuration of the diamond,  replace it by a $^{13}C$ atom, and repeating this in a linear direction, it is possible to have a linear chain of nuclear spins one half and to build a solid state quantum computer. One qubit rotation and controlled-not (CNOT) quantum gates are obtained immediately from this configuration, and CNOT quantum gate is used to determined the design parameters of this quantum computer. 
\end{abstract}
\newpage
\section{\bf Introduction}
\noindent
So far, the idea of having a working quantum computer with enough number of qubits (at least 1000) has faced two main problems: the decoherence \cite{Breuer}-\cite{Lind} due the interaction of the environment with the quantum system, and technological limitations  (pick up signal from NMR quantum computer \cite{Warren} and \cite{Vander}, laser control capability in ion trap quantum computer \cite{Hol} and \cite{MonKim}, physical build up for more than two qubits like in photons cavities \cite{Walther}, atoms traps \cite{Jak} and \cite{Youn}, Josephson's joint ions \cite{Chi}, Aronov-Bhom devices \cite{Kit}, diamond NV device \cite{Child}, or high field and high field gradients in linear chain of paramagnetic atoms with spin one half \cite{Ber1}). In particular, the linear chain of paramagnetic atoms of spin one half became a good mathematical model to make studies of quantum gates \cite{Lo2}, quantum algorithms \cite{Lo3}, and decoherence \cite{Lo4} which could be applied to other to other quantum computers. In this paper, one put together  the ideas of using the diamond stable structure and the linear chain of spin one half nucleus. To do this, on the tetrahedral $^{12}C$ (with nuclear spin zero) configuration of the diamond main structure, one removes one $^{12}C$ element of this configuration an replace it by a $^{13}C$ (with nuclear spin one half) atom, and one repeats this replacement along a linear direction of the crystal. By doing this replacement, one obtains a linear chain of atoms of nuclear spin one half which is protected from the environment by the crystal structure and the electrons cloud. Therefore, one could have a quantum computer highly tolerant to environment interaction and maybe not so difficult to build it,  from the technological point of view. \\Ê\\
\section{\bf $^{12}C$-$^{13}C$ diamond and spin-spin interaction}      
The above idea is represented in Figure 1, where the $^{13}C$ atoms are place on the position of some   $^{12}C$ atoms.  This replacement could be done using the same technics  used to construct the diamond NV structure \cite{Ko}, or using ion implantation technics \cite{Hamm} and neutralization of $^{13}C$ in the diamond \cite{Caza}.  It is assumed in this paper  that this configuration can be built somehow.  
\begin{figure}[t]
\includegraphics[scale=0.5,angle=270]{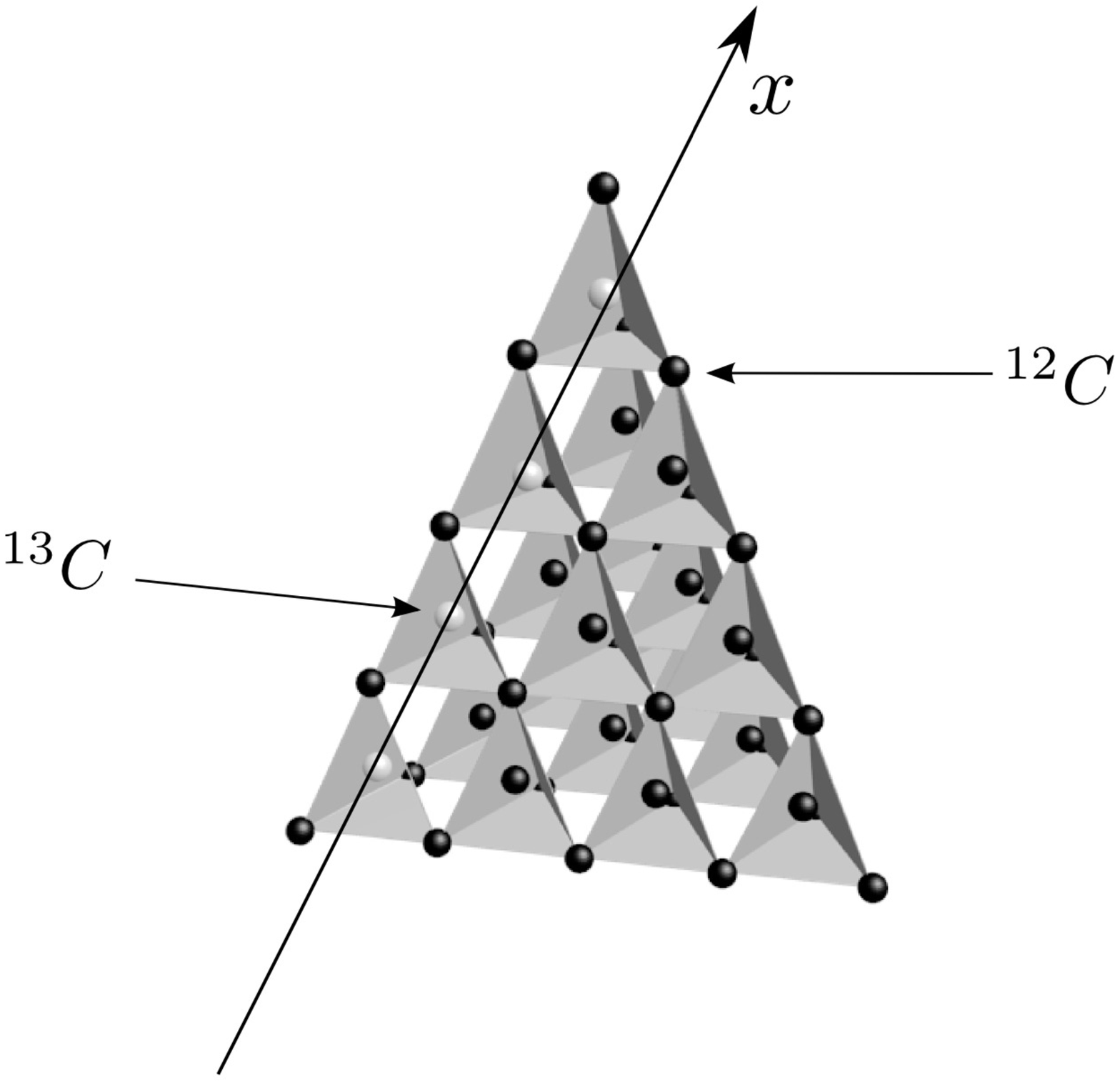}
\centering
    \caption{Diamond $^{12}C$-$^{13}C$. }
    \label{fig:A}
\end{figure}
Now, as one can see, the important interaction on this configuration is the spin-spin interaction between the nucleus of the $^{13}C$ atoms. This interaction is well known \cite{Jack} and is given by
\begin{equation}\label{dd}
U=\frac{\mu_o}{4\pi}\frac{({\bf m_1}\cdot{\bf x})({\bf m_2}\cdot{\bf x})-{\bf m_1}\cdot{\bf m_2}}{|{\bf x}|^3},
\end{equation}
where the magnetic moment ${\bf m_i}, _{i=1,2}$ of  $^{13}C's$ is related with the nuclear spin as
\begin{equation}
{\bf m_i}=\gamma{\bf S_i},
\end{equation}
being $\gamma$ the proton gyromagnetic ratio ($\gamma\approx 2.675\times 10^{8}rad/T\cdot s$). Without loosing the main idea, it will be assumed here that $^{13}C$ magnetic moment is due to proton.  The variable ${\bf x}$ indicates the separation vector between two $^{13}C$ nucleus, which has magnitude $a=|{\bf x}|\sim 10^{-10}m$.  Aligning the chain of $^{13}C$ nucleus along the x-axis of the reference system and assuming Ising interaction between $^{13}C$ nucleus, this energy can be written as
\begin{equation}
U=\frac{J}{\hbar}S_1^zS_2^z,
\end{equation} 
where the coupling constant $J$ has been defined as
\begin{equation}\label{vJ}
J=\frac{\mu_o\gamma^2\hbar}{4\pi a^3}.
\end{equation}
\section{Hamiltonian of the system}
Consider a magnetic field of the form 
\begin{equation}
{\bf B}(x,t)=(b\cos(\omega t+\varphi),-b\sin(\omega t+\varphi), B_0(x)),
\end{equation}
where $b$, $\varphi$, and $\omega$ are the magnitude, the phase, and the frequency of the transverse rf-field. The z-component of the magnetic field has a gradient on the x-axis, determined by the difference on Larmore's frequencies of the $^{13}C's$ nuclear magnetic moments,
\begin{equation}\label{vome}
\left(\frac{\Delta B_0}{\Delta x}\right)=\frac{\Delta\omega}{\gamma\Delta x}.
\end{equation}
The magnetic field at the location of the ith-$^{13}C$ atom is $ {\bf B_i}(t)={\bf B}(x_i,t)$, and the interaction energy of the magnetic moments of the $^{13}C$ atoms with the magnetic field is
\begin{equation}
U=-\sum_{i=1}^N{\bf m_i}\cdot{\bf B_i}(t),
\end{equation}
where $N$ is the number of $^{13}C$ atoms aligned along the x-axis. This energy can be written as
\begin{equation}
U=-\sum_{j=1}^N\omega_jS_j^z-\frac{\Omega}{2}\sum_{k=1}^{N-1}\biggl(e^{i\theta}S_k^{-}+e^{-i\theta}S_k^+\biggr),
\end{equation}
where $\omega_j$ is the Larmore's frequency of the ith-$^{13}C$,
\begin{equation}
\omega_j=\gamma B_{0}(x_j),
\end{equation}
$\Omega$ is the Rabi's frequency,
\begin{equation}
\Omega=\gamma b,
\end{equation} 
$S_j^{-}$ and $S_j^{+}$ are the ascent and descent spin operators,  $S_j^{\pm}=S_j^x\mp iS_j^y$, and $\theta$ has been defined as
\begin{equation}
\theta=\omega t+\varphi.
\end{equation}
Let us consider first and second neighbor interactions among $^{13}C$ nuclear spins, and assuming equidistant separation between any pair of spins, the Hamiltonian of the system is
\begin{eqnarray}
& &H=-\sum_{j=1}^N\omega_jS_j^z+\frac{J}{\hbar}\sum_{k=1}^{N-1}S_j^zS_{j+1}^z+\frac{J'}{\hbar}\sum_{l=1}^{N-2}S_l^zS_{l+2}^z\nonumber\\ \nonumber\\
& &\quad-\frac{\Omega}{2}\sum_{j=1}^N\biggl(e^{i\theta}S_j^{-}+e^{-i\theta}S_j^{+}\biggr),
\end{eqnarray} 
where $J$ is the coupling constant of first neighbor $^{13}C$ atoms, and $J'$ is the coupling constant of second neighbor $^{13}C$ atoms which must be about one order of magnitude lower than $J$. One can write this Hamiltonian as $H=H_0+W(t)$, where $H_0$ and $W$ are defined as
\begin{equation}\label{ho}
H_0=-\sum_{j=1}^N\omega_jS_j^z+\frac{J}{\hbar}\sum_{k=1}^{N-1}S_j^zS_{j+1}^z+\frac{J'}{\hbar}\sum_{l=1}^{N-2}S_l^zS_{l+2}^z,
\end{equation}
and
\begin{equation}\label{wo}
W(t)=-\frac{\Omega}{2}\sum_{j=1}^N\biggl(e^{i\theta}S_j^{-}+e^{-i\theta}S_j^{+}\biggr).Ê
\end{equation}
The operator $H_0$ is diagonal on the basis $\{|\xi\rangle=|\xi_N\dots\xi_1\rangle\}_{\xi_k=0,1}$ of the Hilbert space of $2^N$ dimensionality. Its eigenvalues defines the spectrum of the system,
\begin{equation}\label{en}
E_{\xi}=\frac{\hbar}{2}\left\{-\sum_{j=1}^N(-1)^{\xi_j}\omega_j+\frac{J}{2}\sum_{k=1}^{N-1}(-1)^{\xi_k+\xi_{k+1}}+\frac{J'}{2}\sum_{l=1}^{N-2}(-1)^{\xi_l+\xi_{l+2}}\right\}.
\end{equation}
Since $J'<J\ll\omega_j$ for $_{j=1,\dots,N}$, this spectrum is not degenerated with $E_{|00\dots 0\rangle}$ as the energy of ground state, and $E_{|11\dots 1\rangle}$ as the energy of the most exited state. To calculate the spectrum, one has used the following action of $S_j^z$ operator  
\begin{equation}
S_j^z|\xi\rangle=\frac{\hbar}{2}(-1)^{\xi_j}|\xi\rangle.
\end{equation}
The Schr\"odinger's equation,
\begin{equation}
i\hbar\frac{\partial |\Psi\rangle}{\partial t}=H|\Psi\rangle,
\end{equation}
is solved by proposing a solution of the form
\begin{equation}
|\Psi\rangle=\sum_{\xi}C_{\xi}(t)|\xi\rangle,
\end{equation}
which brings about the following system of first order differential equations on the interaction representation
\begin{equation}\label{as}
i\hbar {\dot a}_{\delta}=\sum_{\xi}a_{\xi} e^{i(E_{\delta}-E_{\xi})t/\hbar}W_{\delta,\xi}(t),
\end{equation}
where $a_{\delta}$ and $W_{\delta,\xi}$ are defined as
\begin{equation}
a_{\delta}(t)=C_{\delta}(t) e^{-iE_{\delta} t/\hbar}
\end{equation}
and
\begin{equation}
W_{\delta, \xi}(t)=\langle\delta|W(t)|\xi\rangle.
\end{equation}
This is very well known procedure to solve time dependent Schr\"odinger's equation, and the solution of Eq. (\ref{as}) brings about he unitary
evolution of the system (given the initial condition $|\Psi_o\rangle$). \\Ê\\
Defining the evolution parameter $\tau$ through the change of variable $t=\omega_o\tau$ ($\omega_o=2\pi MHz$), the parameters $\omega_j$, $\Omega$, $J$ and $J'$ are real numbers given in units of $\omega_o$.   This evolution parameter will be used below in the analysis of the CNOT quantum gate.
\section{ Analysis of the system}
In order to get an operating quantum computer, one  needs to show that, at least, one qubit rotation gate ($N=1$) and two qubits CNOT gate ($N=2$) or three qubits controlled-controlled-not (CCNOT) gate ($N=3$) can be constructed from this quantum system. Because this quantum system is homeomorphic \cite{KF} to the linear chain of paramagnetic atoms with spin one half system \cite{Nie}, it is clear from the point of view of mathematical models that the above gates can be constructed with  this $^{12}C$-$^{13}C$ diamond system. However, one needs to assign realistic workable parameters for the real design of a $^{12}C$-$^{13}C$ diamond quantum computer. To do this, one studies in this section the behavior of a quantum CNOT gate as a function of several parameters.   One neglect one qubit rotation ($N=1, J=J'=0$) because it is obvious that one can get it through an arbitrary pulse on the rf-field with the frequency given by the Larmore's frequency of the qubit ($\omega=\omega_1$), for a single $^{13}C$ atom in the diamond structure. In particular, the NOT quantum gate is obtained using a $\pi$-pulse duration ($\tau=\pi/\Omega$) with this frequency.  Therefore, the study of the CNOT quantum gate is of the most interest ($N=2, J\not=0, J'=0$). Two qubits dynamics is obtained from   
Eqs. (\ref{ho}), (\ref{wo}), and (\ref{as}), resulting the equations
\begin{eqnarray}
& &i\dot a_1=-\frac{\Omega}{2}\bigl(e^{-i(\omega t+\varphi+(E_2-E_1)t/\hbar)}a_2+e^{-i(\omega t+\varphi+(E_3-E_1)t/\hbar)}a_3\bigr)\\ \nonumber\\
& &i\dot a_2=-\frac{\Omega}{2}\bigl(e^{+i(\omega t+\varphi+(E_2-E_1)t/\hbar)}a_1+e^{-i(\omega t+\varphi+(E_4-E_2)t/\hbar)}a_4\bigr)\\ \nonumber\\
& &i\dot a_3=-\frac{\Omega}{2}\bigl(e^{+i(\omega t+\varphi+(E_3-E_1)t/\hbar)}a_1+e^{-i(\omega t+\varphi+(E_4-E_3)t/\hbar)}a_4\bigr)\\ \nonumber\\
& &i\dot a_4=-\frac{\Omega}{2}\bigl(e^{+i(\omega t+\varphi+(E_2-E_1)t/\hbar)}a_2+e^{+i(\omega t+\varphi+(E_4-E_3)t/\hbar)}a_3\bigr)\\ \nonumber
\end{eqnarray}
where the complex variables $a_i$ for $_{i=1,2,3,4}$ correspond to the amplitude of probability to find the system on the states $|00\rangle, |01\rangle, |10\rangle$ and $|11\rangle$. The energies $E_i$ for $_{i=1,2,3,4}$ are deduced from Eq. (\ref{en}). 
Note that $a_1(0)=C_{00}(0)$ and $|a_1(t)|^2=|C_{00}(t)|^2$ (the same for the other variables). CNOT quantum gate corresponds to the transition $|10\rangle \leftrightarrow |11\rangle$, and this one is gotten by selecting the rf-frequency as
\begin{equation}
\omega=\omega_1-J/2.
\end{equation}
Larmore's frequencies are denoted by $\omega_1$ and $\omega_2$,  and $\omega_2$ is parametrized as
\begin{equation}
\omega_2=\omega_1(1+f),
\end{equation}
where $f$ measures the relative change of the frequencies of both qubits. The separation of the $^{13}C$ nucleus, $a$, is parametrized as
\begin{equation}\label{asep}
a=\xi\cdot 10^{-10}m.
\end{equation}
Figure 2 shows the CNOT quantum gate behavior with the initial conditions $C_{00}(0)=C_{01}(0)=C_{11}(0)=0$ and $C_{10}(0)=1$ during a $\pi$-pulse ($\tau=\pi/\Omega$) and with  the parameters
\begin{equation}
B_{01}=0.5~T,\quad \omega_1=21.287,\quad J=0.12,\quad\xi=1,\quad f=0.05. 
\end{equation} 
\begin{figure}[t]
\includegraphics[width=1\textwidth]{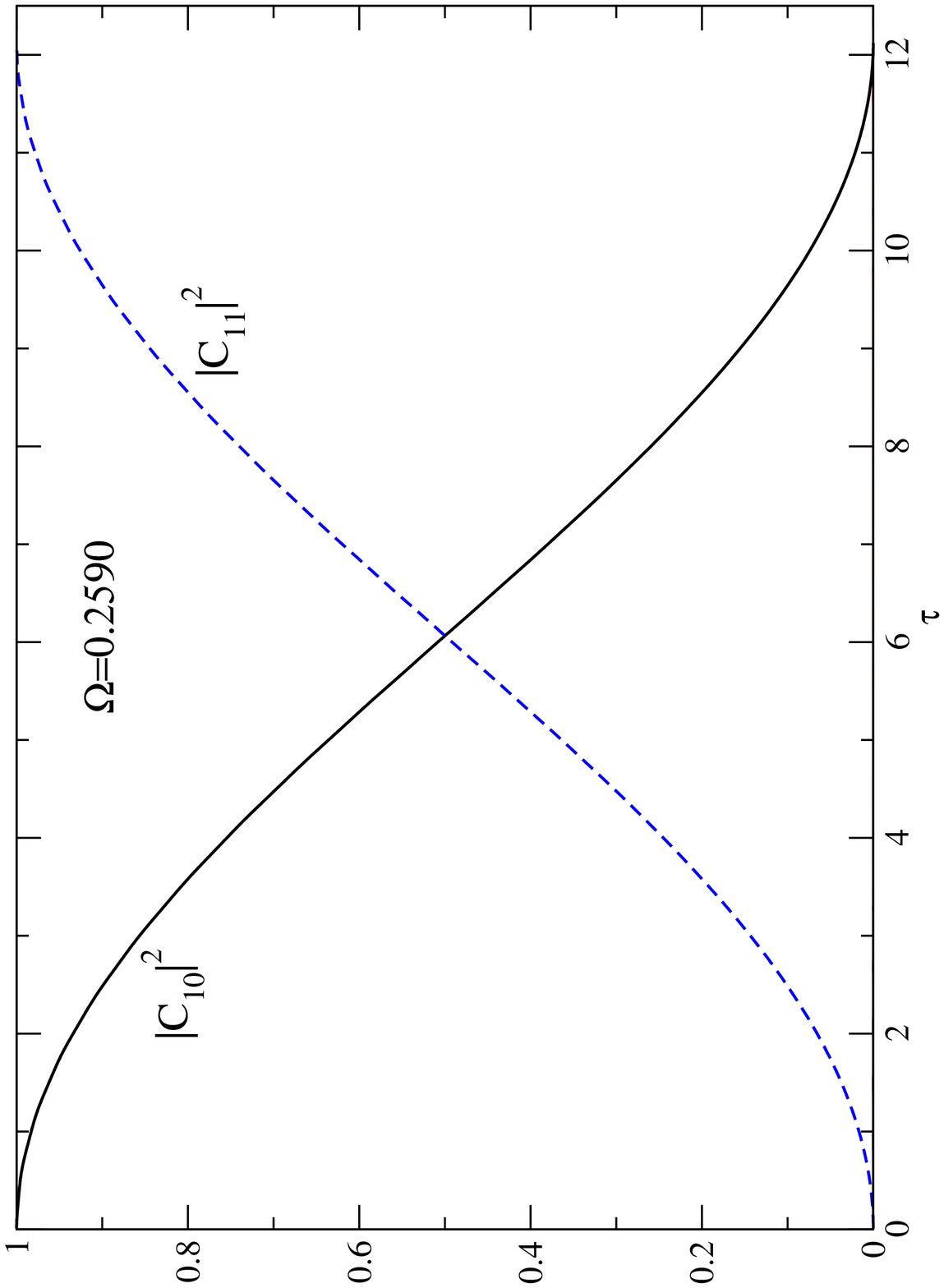}
\centering
    \caption{CNOT quantum gate. }
    \label{fig:B}
\end{figure}
\begin{figure}[t]
\includegraphics[width=1\textwidth]{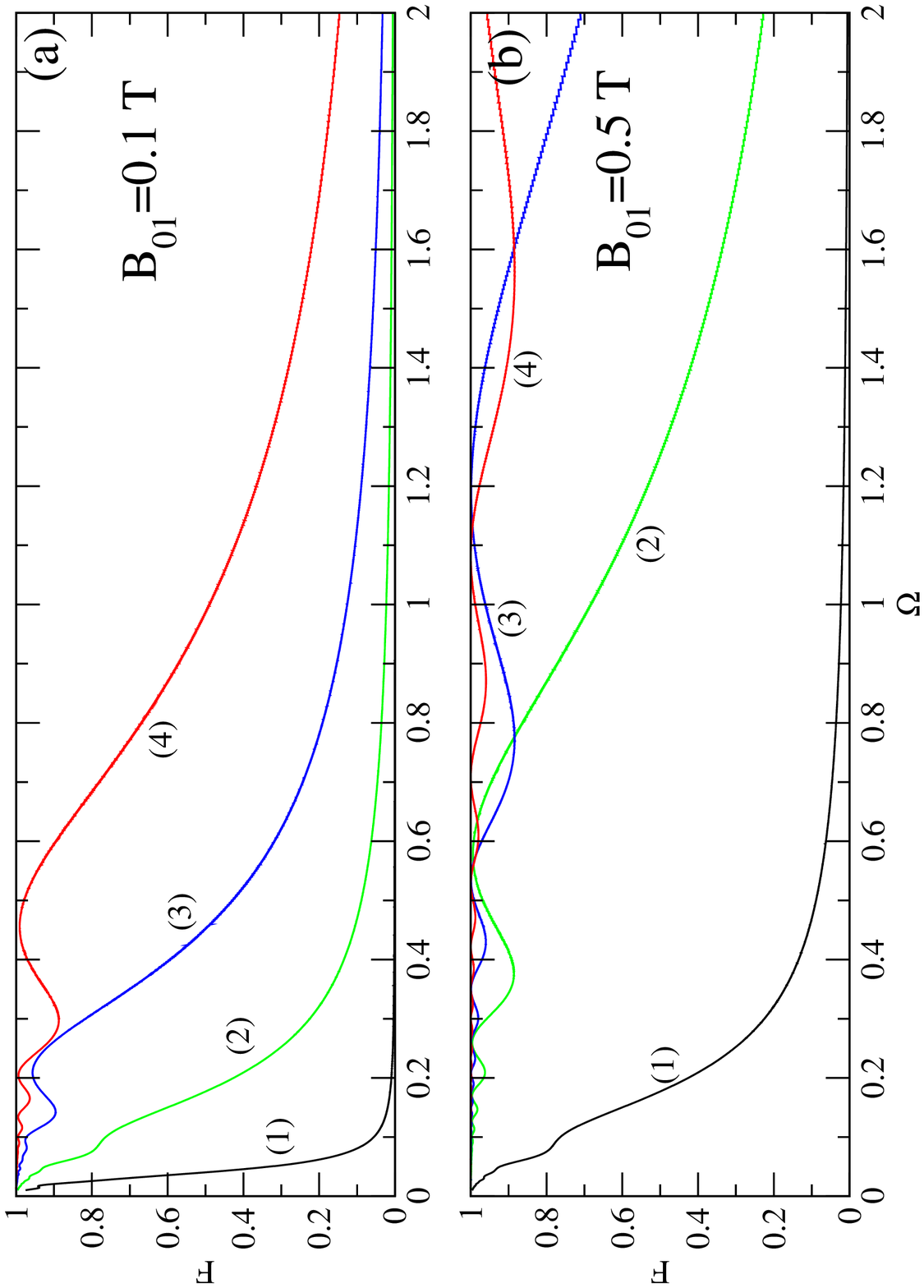}
\centering
    \caption{Fidelity at the end of the $\pi$-pulse.}
    \label{fig:C}
\end{figure}
Figure 3 shows the fidelity parameter,
\begin{equation}
F=|\langle\Psi_{ideal}|\Psi_{real}\rangle|^2,
\end{equation}
at the end of the $\pi$-pulse, as a function of the Rabi's frequency, where $|\Psi_{real}\rangle$ is the state obtained with the simulation, and $|\Psi_{ideal}\rangle$ is the expected state ($|11\rangle$). The simulation was done for two different weak magnetic fields and for  $f=0.01$ (1), $f=0.05$ (2), $f=0.1$ (3), and $f=0.2$ (4).
The oscillations seen on this picture are due to the low and high contribution of the non resonant states ($|00\rangle$ and $|01\rangle$)  to the dynamics of the system, which depends on Rabi's frequency and they are explained by the $2\pi k$-method \cite{Ber1}. As one can see from this picture , the CNOT gate is very well  produced  either with $B_{01}=0.1~T$ and $f=0.2$ or with $B_{01}=0.5~T$ and $f=0.05$. \\ \\
Figure 4 shows the gradient of magnetic field along the x-axis, the coupling constant $J$, and the fidelity $F$ of the CNOT quantum gate as a function of the two qubits separation (characterized by the parameter $\xi$, Eq. (\ref{asep})), having $f=0.05$. As one can see, the fidelity is not sensitive for relatively wide variation of $\xi$, meanwhile  the gradient and coupling constant have the strong variation deduce  from Eq. (\ref{vome}) and Eq. (\ref{vJ}). Considering the separation of the two $^{13}C$ atoms about the the length of the diamond unit cell, one can select $\xi=3$, corresponding to a coupling constant of $J=0.00445$, and a magnetic field gradient of $\Delta B_0/a\approx 0.83\times 10^6 T/m$.
\begin{figure}[t]
\includegraphics[width=1\textwidth]{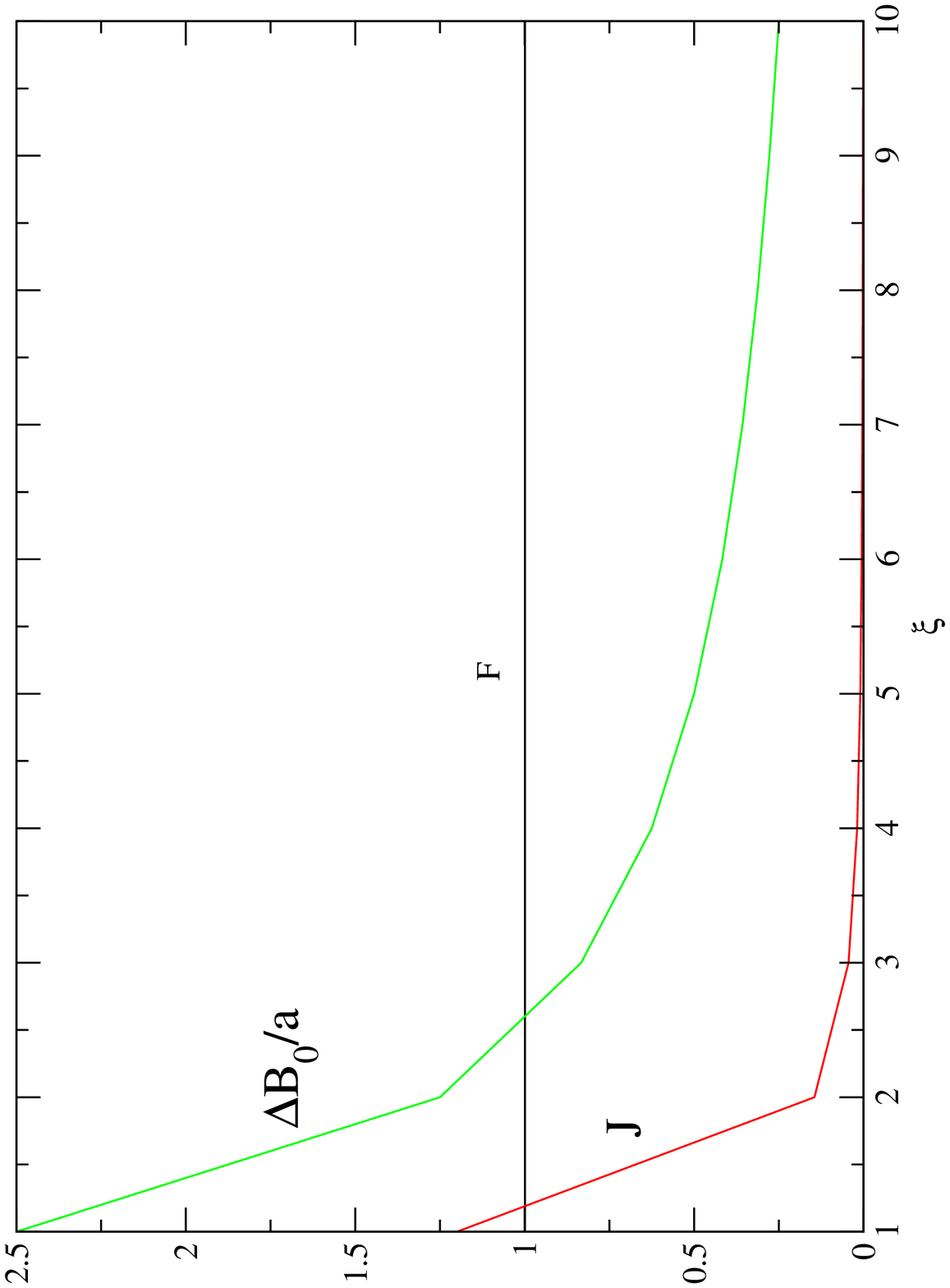}
\centering
    \caption{Effect of $^{13}C$-$^{13}C$ separation.  }
    \label{fig:D}
\end{figure}
One needs to mention that in the case the alignment of the $^{13}C$ atoms be along the z-axis (the same direction of the longitudinal magnetic field), the coupling constant deduced from Eq. (\ref{dd}) would be given by  $-2J$, with $J$ given by Eq. (\ref{vJ}), and basically the results are the same as  the presented here. 
\newpage\noindent
 According to these results, one has now an  idea of the value of the parameters for the design of a quantum computer with the $^{12}C$-$^{13}C$ diamond quantum system: (a) Separation between $^{13}C$ atoms is $a=3\times 10^{-10}m$ which can be aligned along the x-axis, (b) coupling constant is $J=0.00445 (2\pi~MHz)$, (c) longitudinal magnetic field is $B_{01}=0.05~T$, (d) gradient of this longitudinal magnetic field along the x-axis is $\Delta B_0/a=0.83\times 10^6T/m$, and (e) magnitude of the rf-magnetic field on the plane x-y is $b=0.00608~T$ (Rabi's frequency $\Omega=0.259 (2\pi~Mhz)$). \\Ê\\
Although the gradient of the magnetic field might be a concern, the magnitude of the longitudinal magnetic field is low enough to think that this gradient can be achieved.  The scalability of the system is clear, the read out system could be based on single spin measurement technics \cite{Ru}, and studies on decoherence remains to be done on this system. This quantum computer resembles  a solid state NMR system \cite{Duer}.
\section{ Conclusion and discussion}
It was shown that by removing a $^{12}C$ atom,  replace it by a $^{13}C$ atom in the tetrahedral configuration of the diamond, and doing this process periodically in a linear direction, one could get a linear chain of nuclear spins one half which can be work as a quantum computer. The interaction between $^{13}C$ atoms is governed by the magnetic dipole-dipole interaction, and the parameters of a possible quantum computer design were determined by studying the quantum CNOT gate with two qubits. Although there might be a concern about the gradient of the magnetic field along the lines of $^{13}C$ atoms, it must not be so difficult to get this gradient since the magnitude of this magnetic field is relatively low (0.5 T).  In principle, it is possible to replace a $^{12}C$ atom by any other spin one half atom. However, an unclose configuration of electrons in the lattice makes  necessarily to take into account the interaction of electrons with this atom ( as it is the case of diamond NV configuration) which makes the analysis and the quantum computer much more complicated and sensitive to environment  interaction. The misplacement of the $^{13}C$ atom along the x-axis produces different coupling constant in the interaction, but according to Figure 4, the fidelity of the CNOT quantum gate does not change, and one would expect  the same result for quantum algorithms. The displacement of $^{13}C$ atoms off x-axis changes the coupling constant and the interaction itself, which has to be studied. In addition,  it still remains to study the  decoherence on quantum gates and quantum algorithms of system. 

\newpage


\begin{thebibliography}{10}

\bibitem{Breuer} 
H. -P. Breuer and F. Petruccione,
\newblock{"The Theory of Open Quantum Systems,"}
\newblock{ Oxford University Press, 2006.}

\bibitem{CaLeg}
A.O. Caldeira and  A.T. Legget, 
\newblock {\em  Physica A},  {\bf 121}, 587 (1983).

\bibitem{UnZu}
W.G. Unruh and W.H. Zurek, 
\newblock {\em  Phys. Rev. D} {\bf 40}  1071, (1989).

\bibitem{Paz1}
B.L. Hu, J.P. Paz, and Y. Zhang, 
\newblock {\em  Phys. Rev. D} {\bf 45}, 2843 (1992).

\bibitem{Ven1}
A. Venugopalan, 
\newblock {\em  Phys. Rev. A} {\bf 56}, 4307 (1997).

\bibitem{Zeh}
H.D. Zeh,  
\newblock {\em  Found. Phys.} {\bf 3}, 109 (1973).

\bibitem{PazZu}
J.P. Paz and W.H. Zurek, 
\newblock {\em  Proc. Les Houches, 111A}, 409  (1997).

\bibitem{Lind}
G. Lindblad, 
\newblock {\em  Commun. Math. Phys.,} {\bf 48}, 119 (1976).

\bibitem{Warren}
W.S. Warren, 
\newblock{\it The usefulness of NMR quantum computing},
\newblock{Sciencie, {\bf 277}, 1688 (1997).}

\bibitem{Vander}
L.M.L. Vandersypen, M. Steffen, G. Breyta, C.S. Yannoni, M.H. Sherwood, and I.L. Chuang,
\newblock{Nature, {\bf 414}, 883 (2001).}

\bibitem{Hol}
M.H. Holzschelter,
\newblock{ Los Alamos Science, {\bf 27}, 264 (2002).}

\bibitem{MonKim}
C. Monroe and J. Kim,
\newblock{Science, {\bf 339}, 1164 (2013).}

\bibitem{Walther}
H. Walter, B.T.H. Varcoe,  B.G. Englert and T. Becker, 
\newblock{Rep. Prog. Phys. {\bf 69}, 1325 (2006).}

\bibitem{Jak}
D. Jaksch, J.I. Cirac, P. Zoller, S.L. Rolston, R. Cot\'e and M.D. Lukin
\newblock{Phys. Rev. Lett., {\bf 85}, 2208 (2000).}

\bibitem{Youn}
K.C. Younge, B. Knuffman, S.E. Anderson and G. Raithel,
\newblock{ Phys. Rev. Lett., {\bf 104}, 173001 (2010).}

\bibitem{Chi}
I. Chiorescu, Y. Nakamura, C.J.P.M. Harmans and J.E. Mooij,
\newblock{Science, {\bf 299}, 1869 (2003).}

\bibitem{Kit}
A. Yu. Kitaev, 
\newblock{Annals Phys. {\bf 303}, 2 (2003).}

\bibitem{Child}
L. Childress and R. Hanson,
\newblock{MRS Bulletin, {\bf 38}, 134 (2013).}

\bibitem{Ber1}
G. P. Berman, D.I. Kamenev, D.D. Doolen, G.V. L\' opez  and V.I. Tsifrinovich, 
\newblock {\em Contemp. Math., {\bf 305} 13 (2002).}

\bibitem{Lo2}
G.V. L\'opez and L.Lara,
\newblock {\em J.  Phys. B: At. Mol.  Opt. Phys.,} {\bf 39}, 3897 (2006).

\bibitem{Lo3}
G.V. L\'opez, T. Gorin, and L.Lara,
\newblock {\em J. Phys. B: At. Mol. Opt. Phys.,} {\bf 41}, 055504 (2008).

\bibitem{Lo4}
G.V. L\'opez and P. L\'opez,
\newblock{J. Mod. Phys., {\bf 3}, 85 (2012).}

\bibitem{Hamm}
R.W. Hamm, M.E: Hamm,
\newblock{\it Industrial Acceleretors and Their Applications},
\newblock{World Scientific, ISBN 978-981-4307-04-8, (2012).}

\bibitem{Caza}
M.A. Cazalilla, N. Lorente, R.D.Mui\~no, J.P.Gauyacq, D. Teillet-Billy and P.M. Echenique,
\newblock{Phys. Rev. B, {\bf 58},  13991 (1998).}

\bibitem{Ko}
K. Lakoubovskii and G.J. Adriaenssens,
\newblock{J. Phys. Condens.Matter, {\bf 13}, 6015 (2001).}

\bibitem{Jack}
J.D. Jackson,
\newblock{\it Classical Electrodynamics, Third edition}, chapter 5.6, John Wiley and Sons, Inc. (1999).

\bibitem{Nie}
M. A. Nielsen and I.L. Chuang,
\newblock{\it Quantum Computation and Quantum Information}, Cambridge University Press, (2000).

\bibitem{Ru}
D. Rugar, R. Budakian, H.J. Mamin and B.W. Chui,
\newblock{Nature, {\bf 430}, 329 (2004)}

\bibitem{Duer}
M.J. Duer,
\newblock{\it Introduction to Solid-State NMR Spectroscopy}, Blackwell, Oxford (2004).

\bibitem{KF}
A.N. Kolmogorov and S.V. Fomin,
\newblock{\it Introductory Real Analysis}, Dover Publications, Inc.  (1970).


\end{thebibliography}
\end{document}